\documentclass{emulateapj}
\usepackage{apjfonts}

\def\ltsima{$\; \buildrel < \over \sim \;$}
\def\simlt{\lower.5ex\hbox{\ltsima}}
\def\gtsima{$\; \buildrel > \over \sim \;$}
\def\simgt{\lower.5ex\hbox{\gtsima}}

\newcommand{\suzaku}{{\it Suzaku}}
\newcommand{\Suzaku}{{\it SUZAKU}}
\newcommand{\swsix}{{Swift J0601.9--8636}}
\newcommand{\Swsix}{{SWIFT J0601.9--8636}}
\newcommand{\swone}{{Swift J0138.6--4001}}

\newcommand{\swift}{{\it Swift}}
\newcommand{\integral}{{\it INTEGRAL}}

\newcommand{\rosat}{{\it ROSAT}}

\newcommand{\etal}{{\it et al.}}

\newcommand{\ergs}{erg cm$^{-2}$ s$^{-1}$}
\newcommand{\erg}{erg s$^{-1}$}

\newcommand{\nh}{$N_{\rm H}$}
\newcommand{\cosmo}{($H_0$, $\Omega_{\rm m}$, $\Omega_{\lambda}$)}

\lefthead{Ueda et al.}
\righthead{Suzaku Observations of Swift AGNs}
\begin{document}

\title{{\it Suzaku} Observations of Active Galactic Nuclei Detected in the
{\it Swift}/BAT Survey: \\
Discovery of ``New Type'' of Buried Supermassive Black Holes}

\author{
Yoshihiro Ueda\altaffilmark{1},
Satoshi Eguchi\altaffilmark{1},
Yuichi Terashima\altaffilmark{2},
Richard Mushotzky\altaffilmark{3},
Jack Tueller\altaffilmark{3},
Craig Markwardt\altaffilmark{3},
Neil Gehrels\altaffilmark{3},
Yasuhiro Hashimoto\altaffilmark{4},
Stephen Potter\altaffilmark{4}
}

\altaffiltext{1}{Department of Astronomy, Kyoto University, Kyoto
606-8502, Japan}
\altaffiltext{2}{Department of Physics, Faculty of Science, Ehime
University, Matsuyama 790-8577, Japan}
\altaffiltext{3}{NASA/Goddard Space Flight Center, Greenbelt, MD 20771, USA}
\altaffiltext{4}{SAAO, P.O. Box 9, Observatory 7935, South Africa}

\begin{abstract}

We present the \suzaku\ broad band observations of two AGNs detected
by the \swift /BAT hard X-ray ($>$15 keV) survey that did not have
previous X-ray data, \swsix\ and \swone. The \suzaku\ spectra reveals
in both objects a heavily absorbed power law component with a column
density of \nh\ $\simeq 10^{23.5-24} {\rm cm}^{-2}$ that dominates
above 10 keV, and an intense reflection component with a solid angle
$\simgt 2\pi$ from a cold, optically thick medium. We find that these
AGNs have an extremely small fraction of scattered light from the
nucleus, $\simlt 0.5\%$ with respect to the intrinsic power law
component. This indicates that they are buried in a very
geometrically-thick torus with a small opening angle and/or have
unusually small amount of gas responsible for scattering. In the
former case, the geometry of \swsix\ should be nearly face-on as
inferred from the small absorption for the reflection component. The
discovery of two such objects in this small sample implies that there
must be a significant number of yet unrecognized, very Compton thick
AGNs viewed at larger inclination angles in the local universe, which
are difficult to detect even in the currently most sensitive optical
or hard X-ray surveys.

\end{abstract}


\section{Introduction}

Many observations imply the presence of a large number of heavily obscured
Active Galactic Nuclei (AGNs) in the local universe
\citep[e.g.,][]{mai98,ris99}. The number density of AGNs subject to
absorption with a line-of-sight Hydrogen column density log \nh\
$\simgt$ 23.5 cm$^{-2}$ is a key parameter in understanding the
accretion history of the universe
\citep[e.g.,][]{set89,fab99}. According to synthesis models of the
X-ray background (XRB) \citep[e.g.,][]{ued03,gil07}, much of the
peak intensity of the XRB at 30 keV should be produced by these
objects. In spite of the potential importance of their contribution to
the growth of supermassive black holes \citep{mar04}, the nature of
this population of AGNs, even in the local universe, is only poorly
understood due to strong biases against detecting them. In these
objects, the direct emission in the UV, optical, and near IR bands as
well as at $E<10$ keV from the nucleus is almost completely blocked by
obscuring matter, making it difficult to probe the central engine.

Hard X-ray surveys at energies above 10--15 keV provide us with an
ideal opportunity to select this population of AGNs as long as the
column density is less than log \nh\ $\approx 24.5$ cm$^{-2}$. Recent
surveys performed with \swift /BAT (15--200 keV; \citealt{mar05}) and
\integral\ (10--100 keV; \citealt{bas06,bec06,saz07}), because of
their relative insensitivity to absorption, are providing one of the
most unbiased AGN samples in the local universe including Compton
thick AGNs, i.e., those with log \nh\ $>$ 24 cm$^{-2}$. In fact, these
surveys have started to detect hidden AGNs in the local universe
located in galaxies that were previously unrecognized to contain an
active nucleus at other wavelengths.

To unveil the nature of these new hard X-ray sources, follow-up
observations covering a broad energy band are crucial. In this paper,
we present the first results from follow-up observations with the
\suzaku\ observatory \citep{mit07} of the AGNs \swsix\ and \swone\
detected in the \swift /BAT survey. These targets are essentially
randomly selected from a bright \swift\ AGN sample for which soft
X-ray ($<$10 keV) spectroscopic observations had never been performed,
and are thus reasonable representatives of unknown AGN populations
selected by hard X-rays above 15 keV. In \S~3, we also present an
optical spectrum of \swsix\ taken at the South African Astronomical
Observatory (SAAO) 1.9-m telescope. The cosmological parameters
\cosmo\ = (70 km s$^{-1}$ Mpc$^{-1}$, 0.3, 0.7) \citep{spe03} are
adopted throughout the paper.

\section{The \Suzaku\ Observations and Results}

Table~1 summarizes our targets and observation log. \swsix\ is
optically identified as the galaxy ESO 005--G004 \citep{lau82} at
$z=0.0062$ with no previous firm evidence for AGN activity. The
optical counterpart of \swone\ is the galaxy ESO 297--G018
\citep{lau82} at $z=0.0252$, which was identified as a narrow line AGN
\citep{kir90}.

\suzaku, the 5th Japanese X-ray satellite, carries four sets of X-ray
mirrors each with a focal plane X-ray CCD camera, the X-ray Imaging
Spectrometer (XIS-0, XIS-1, XIS-2, and XIS-3; \citealt{koy07}), and a
non-imaging instrument called the Hard X-ray Detector (HXD;
\citealt{tak07}), which consists of the Si PIN photo-diodes and GSO
scintillation counters. The XIS and PIN simultaneously covers the
energy band of 0.2--12 keV and 10--70 keV, respectively. The unique
capabilities of \suzaku , high sensitivity in the 12--70 keV band and
broad band coverage with good spectral resolution, are critical for
studies of highly absorbed AGNs.

We observed \swsix\ and \swone\ with \suzaku\ on 2006 April 14 and
June 5 for a net exposure of 20 and 21 ksec (for the XIS),
respectively. Standard analysis was made on data products, which were
processed with the latest calibration (version 1.2). We detected both
sources at high significance with the XIS and PIN. The XIS
spectra were accumulated within a radius of 2 arcmin
around the detected position. The background was taken from a source
free region in the field of view. The spectra of three Front-side
Illuminated CCD (hereafter FI-XIS; XIS-0, 2, and 3) are summed
together, while that of Back-side Illuminated CCD (BI-XIS; XIS-1) is
treated separately in the spectral fit. Examining the spectra of
$^{55}$Fe calibration source, we verify that the energy scale and
resolution is accurate better than 10 eV and 60 eV levels,
respectively, at 5.9 keV. For the analysis of the PIN, we
utilized only data of Well units with the bias voltage set at 500 eV
(W0,1,2,3 for \swsix\ and W1,2,3 for \swone) with the best available
models of PIN background provided by the HXD team\footnote{ver1.2\_d
for \swsix\ and ver1.2\_w123 for \swone}.

\tabcolsep=2pt
\tablefontsize{\scriptsize}
\begin{deluxetable}{cccccc}
\tablenum{1}
\tablecaption{Targets and Observation Log\label{tbl-1}}
   \tablehead{\colhead{Swift} &\colhead{Optical ID} &\colhead{redshift}
&\colhead{Start Time (UT)} &\colhead{End Time}
&\colhead{Exposure\tablenotemark{a}}}
\startdata
J0601.9--8636 & ESO 005--G004 &0.0062&2006/04/13 16:24 & 04/14 01:52
& 19.8 ksec\\
J0138.6--4001 & ESO 297--G018 &0.0252&2006/06/04 18:13 & 06/05 05:00
& 21.2 ksec\\
\enddata
\tablenotetext{a}{Based on a good time interval for the XIS-0.}
\end{deluxetable}

To obtain the best constraint from the entire data, we perform a
simultaneous fit to the spectra of XIS (FI and BI), PIN, and the 
archival Swift BAT,
which covers the 0.2--200 keV band as a whole. The BAT spectra consist
of four energy bins over the 15--200 keV range and are useful to
constrain the power law index. Here we allow the relative flux
normalization between \suzaku\ and \swift\ (BAT) to be a free
parameter, considering possible time variability between the
observations. We fixed the normalization ratio between the FI-XIS and
the PIN based on the calibration result using the Crab Nebula. All the
absolute fluxes quoted in this paper refer to the flux calibration of
the FI-XIS. We find that the 10--50 keV PIN fluxes of \swsix\ and
\swone\ are $1\times10^{-11}$ and $4\times10^{-11}$ \ergs , indicating
time variability by a factor of 0.5 and 1.6, respectively, compared
with the averaged flux measured by the \swift /BAT over the past 9
months (Tueller \etal , in preparation).

Figure~1 shows the FI-XIS and PIN spectra unfolded for the detector
response (for clarity the BI-XIS and BAT spectra are not plotted). The
X-ray spectrum of \swsix\ below 10 keV is dominated by a hard
continuum with few photons below 2 keV, consistent with the previous
non-detection in soft X-rays (an upper limit of $1.3\times10^{-13}$
\ergs\ in the 0.1--2.4 keV band by the \rosat\ All Sky Survey;
\citealt{vog00}). We find that the broad band spectrum can be well
reproduced with a model consisting of a heavily absorbed power law
with log \nh\ $\simeq 24$ cm$^{-2}$, which dominates above 10 keV, and
a mildly absorbed reflection component from cold matter accompanied by
a narrow fluorescence iron-K line, which dominates below 10 keV. The
large column density is consistent with the observed equivalent width
(EW) of the iron-K line, $\approx$ 1 keV \citep{lev02}. \swone\ shows
a similar spectrum but with a smaller absorption of log \nh\ $=23.7$
cm$^{-2}$ for both transmitted and reflected components.

\begin{figure}
\epsscale{1.0}
\plotone{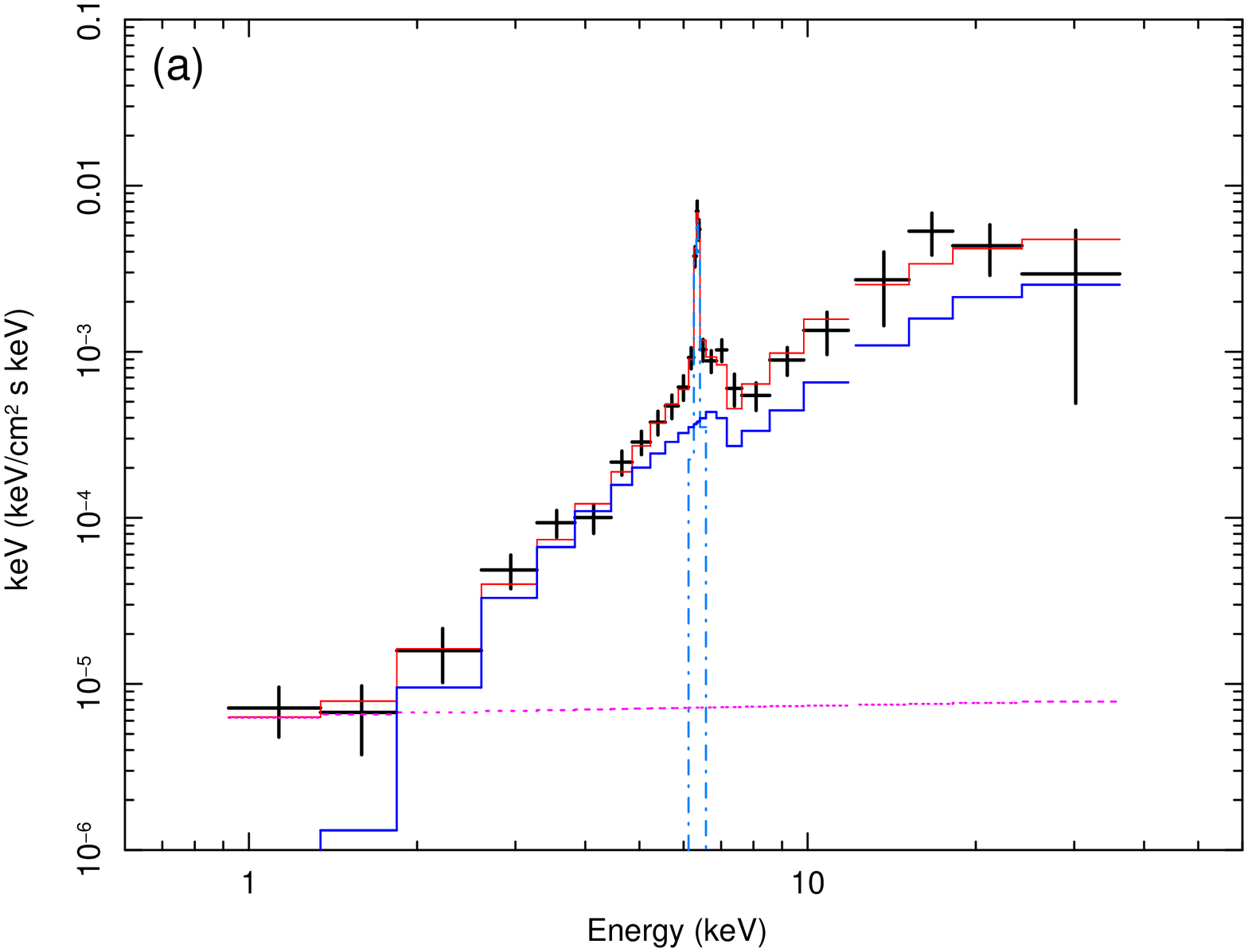}
\plotone{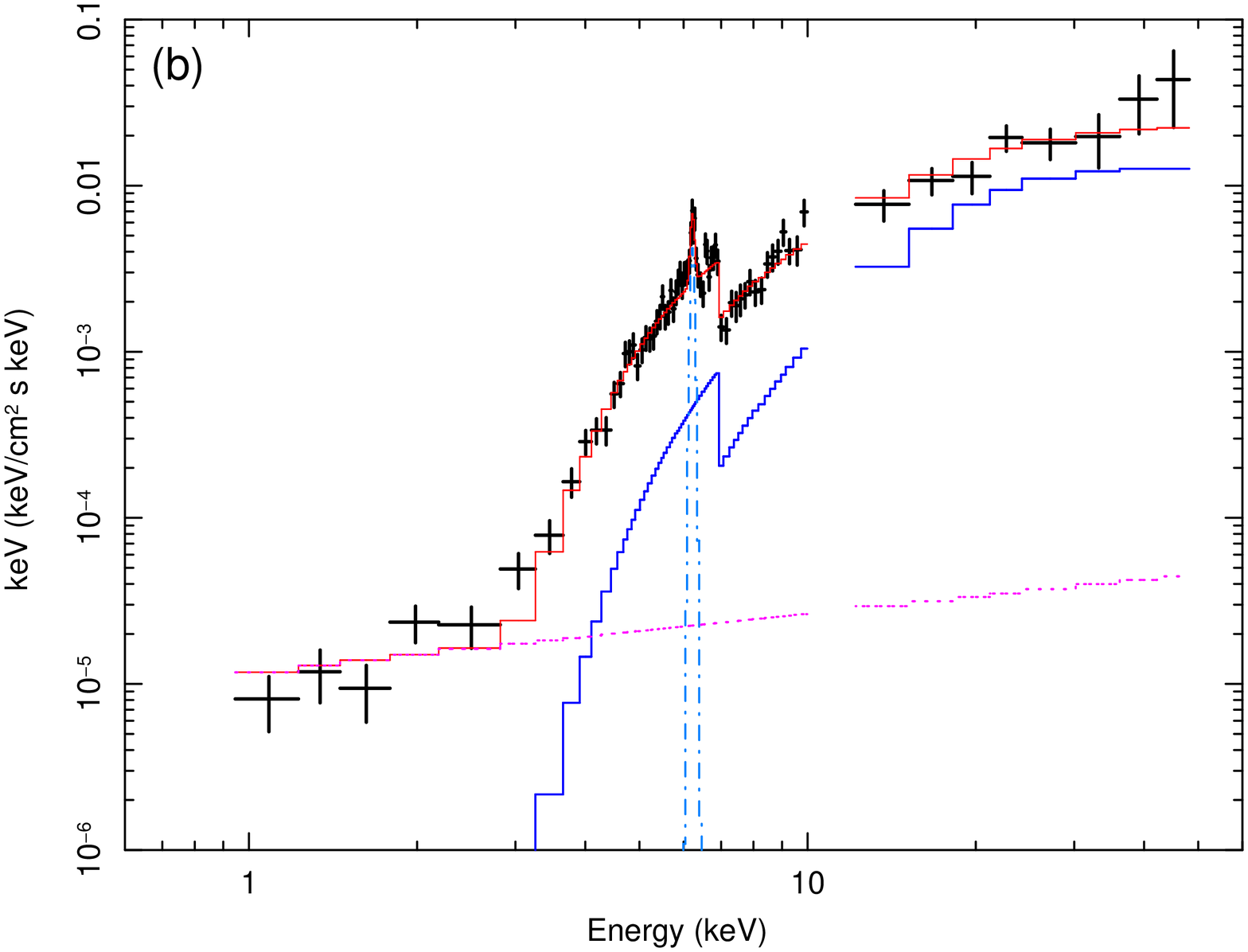}
\caption{ The broad-band energy spectra of (a) \swsix\ and (b) \swone\
unfolded for the detector response in units of $E^2 F(E)$, where
$F(E)$ is the photon spectrum. For clarity, we only plot the summed
spectrum of the three FI-XIS (below 12 keV), and those of PIN
(above 12 keV), while the spectral fit is performed to the whole
XIS+PIN+BAT data including the BI-XIS. The crosses (black) represent the
data with $1\sigma$ statistical errors. The histograms show the
best-fit model with separate components. 
The upper solid line (red), dot-dashed line (light blue), lower
solid line (blue), and dotted line (magenta) correspond to the total,
iron-K emission line, reflection component, and scattered component,
respectively.
\label{fig1}}
\end{figure}

The spectral model is represented as
$$
F(E) = e^{-\sigma (E) N_{\rm H}^{\rm Gal}} [
f \; A E^{-\Gamma}
+ e^{-\sigma (E) N_{\rm H}} A E^{-\Gamma}
+ e^{-\sigma (E) N_{\rm H}^{\rm refl}} C(E)
+ G(E) ]
,
$$ where $N_{\rm H}^{\rm Gal}$ is the Galactic absorption column
density fixed at $2.0\times10^{20} {\rm cm}^{-2}$ for both targets
\citep{dic90}, $N_{\rm H}$ the local absorption column density at the
source redshift for the transmitted component, $N_{\rm H}^{\rm refl}$
that for the reflected component (assumed to be the same as $N_{\rm
H}$ for \swone ), and $\sigma (E)$ the cross section of photo-electric
absorption. The term $C(E)$ represents the reflection component,
calculated using the code in \citet{mag95}; we leave the solid angle $\Omega$
of the reflector as a free parameter by fixing the inclination angle
at 60$^{\circ}$ and cutoff energy at 300 keV, assuming Solar
abundances for all elements. $R(\equiv\Omega/2\pi) >1$ means 
the transmission efficiency should be $\simlt 1/R$ (see \S~4).
The term $G(E)$ is the narrow iron-K emission line modelled by a
Gaussian profile, where we fix the $1\sigma$ width at 50 eV to take
account of the response uncertainty (and hence, the line should be
considered to be unresolved). The best-fit parameters are summarized
in Table~2 with the intrinsic 2--10 keV luminosity, $L_{\rm 2-10}$, 
corrected for the absorption and transmission efficiency of $1/R$.

\tabcolsep=6pt
\tablefontsize{\footnotesize}
\begin{deluxetable}{cccc}
\tablenum{2}
\tablecaption{Best Fit Spectral Parameters\label{tbl-2}}
\tablehead{\colhead{} &\colhead{Swift} &\colhead{J0601.9--8636}
&\colhead{J0138.6--4001} }
\startdata
(1) & \nh\ ($10^{22} {\rm cm}^{-2}$) & $101^{+54}_{-38}$& $46\pm4$\\
(2) & $\Gamma$          & $1.95^{+0.36}_{-0.33}$ & $1.66^{+0.16}_{-0.04}$\\
(3) &$R$       & $1.7^{+3.5}_{-0.9}$ & $2.1^{+0.4}_{-1.2}$\\
(4) &$N_{\rm H}^{\rm refl}$ ($10^{22} {\rm cm}^{-2}$) & 
$2.9^{+5.3}_{-1.4}$& (= \nh ) \\
(5) &$f_{\rm scat}$ (\%) & $0.20\pm0.11$ & $0.23^{+0.23}_{-0.16}$\\
(6) &$E_{\rm cen}$ (keV)& $6.38\pm0.02$& $6.38\pm0.03$\\
(7) &E.W. (keV)& $1.06\pm0.16$& $0.20\pm0.05$\\
(8) &$F_{\rm 2-10} $ (\ergs ) & $1.1\times10^{-12}$&$3.3\times10^{-12}$\\
(9) &$F_{\rm 10-50}$ (\ergs ) & $9.8\times10^{-12}$&$3.9\times10^{-11}$\\
(10) &$L_{\rm 2-10} $ (\erg ) & $8.3\times10^{41}$&$3.9\times10^{43}$\\
      &$\chi^2$(dof) & 20.0 (27)& 101.3 (87)\\
\enddata
\tablecomments{
(1) The line-of-sight hydrogen column density for the transmitted 
component; (2) The power law photon index;
(3) The relative strength of the reflection component to the
transmitted one, defined as $R\equiv\Omega/2\pi$, where $\Omega$ is
the solid angle of the reflector viewed from the nucleus;
(4) The line-of-sight hydrogen column density for the refection 
component (assumed to be the same as \nh\ for \swone );
(5) The fraction of a scattered component relative to the intrinsic power
law corrected for the transmission efficiency of $1/R$ when $R>1$; (6)
The center energy of an iron-K emission line at rest frame. The
1$\sigma$ line width is fixed at 50 eV; (7) The observed equivalent
width of the iron-K line with respect to the whole continuum; (8)(9)
Observed fluxes in the 2--10 keV and 10--50 keV band; (10) The 2--10
keV intrinsic luminosity corrected for the absorption and transmission
efficiency of $1/R$.\\
The errors are 90\% confidence limits for a single parameter.}
\end{deluxetable}

We confirm that these results are robust, within the statistical
error, given the systematic errors in the background estimation of the
PIN detector \citep{kok07}. In the case of \swone, where photon
statistics is dominated by the XIS data, we limit the allowed range of
photon index to $\Gamma=1.63-2.02$ in the simultaneous fit, being
constrained from the BAT spectrum. We have limited 
$R<2.5$ for this source to avoid physical
inconsistency between $R$ and the EW of an iron-K line; otherwise
(i.e., in more ``reflection-dominated'' spectra), we should expect a
larger EW than the observed value of $\approx$ 0.2 keV.

\section{Optical Spectrum of \Swsix }

We performed an optical spectroscopic observation of \swsix\ (ESO
005--G004) during the night of 2007 March 16, using the SAAO 1.9-m
telescope with the Cassegrain spectrograph. Grating six, with a
spectral range of about 3500-5300 \AA\ at a resolution around 4 \AA,
was used with a 2 arcsec slit placed on the center of the galaxy for a
total integration time of 2400 s. To derive the sensitivity curve, we
fit the observed spectral energy distribution of standard stars with
low-order polynomial. The co-added, flux-calibrated spectrum in the
4000-5500 \AA\ range is shown in Figure~2. It reveals a rather
featureless spectrum with no evidence for H$\beta$ or [O~III]
$\lambda$5007 emission lines, typical for this type of non-active
edge-on galaxy within this spectral range. The 90\% upper limit on the
[O~III] flux is conservatively estimated to be $3\times10^{-15}$ \ergs
, corresponding to a luminosity of $3\times10^{38}$ \erg . This yields
the ratio of the intrinsic 2--10 keV luminosity to the {\it observed}
[O~III] luminosity of $> 2800$. Although the (unknown) extinction
correction for [O~III] could reduce the value, the result is
consistent with \swsix\ having an intrinsically weak [O~III] emission
relative to hard X-rays compared with other Seyfert galaxies
\citep{bas99,hec05,net07}.  In particular, this object would not have
been selected to be an AGN on the basis of its [O~III] or H$\beta$
emission.

\begin{figure}
\epsscale{1.0}
\plotone{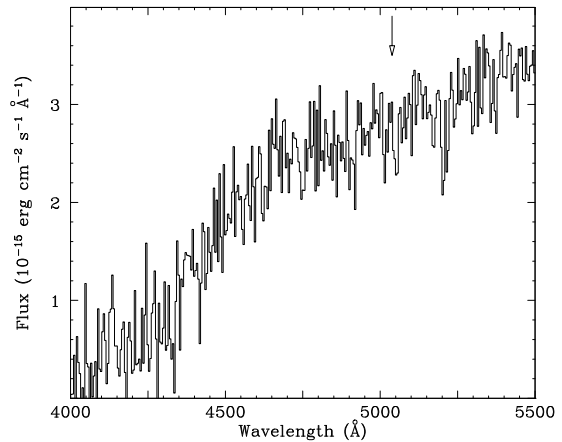}
\caption{
The optical spectrum of the nucleus region of \swsix\
  (ESO 005--G004) in the 4000-5500 \AA\ wavelength range, taken with
  the SAAO 1.9-m telescope. The arrow denotes the position of the
  [O~III] $\lambda5007$ line.
\label{fig1}}
\end{figure}

\section{Discussion}

Both sources show an intense reflection component relative to the
transmitted one. Using the standard reflection model \citep{mag95}, we
find that the solid angle of the reflector $\Omega/2\pi$ viewed from
the nucleus exceeds unity, which is apparently unphysical if
attributed only to geometry. This implies that a part of the direct
emission is completely blocked by non-uniform material in the line
of sight even above 10 keV. The reflection-dominated nature of the
spectra of heavily obscured AGNs, if common, has an impact on the
population synthesis model of the XRB, where a much weaker reflection
is assumed for type 2 AGNs \citep{gil07}. Another possibility is that this
apparent very high reflection fraction is due to time
variability, that is, the decrease of the flux in the transmitted
light is echoed with a time delay corresponding to the difference in
light paths between the emitter, reflector, and observers.

It is remarkable that both \swsix\ and \swone\ have a very small
amount of soft X-ray scattered emission, less than 0.46\% of the
intrinsic power law component. (If we fix $N_{\rm H}^{\rm refl} = 0$
in the spectral fit of \swsix , then we obtain a photon index of
$1.80\pm0.29$ and no significant scattered component with a 90\% upper
limit of 0.47\%.) As far as we know these are amongst the lowest
scattered fractions ever seen from an absorbed AGN
\citep{tur97,cap06}. In optically selected Seyfert 2 galaxies, the
presence of prominent soft X-ray emission is common
\citep[e.g.,][]{gua05}. Such emission probably originates from the
same extended gas responsible for the optical [O~III] emission
\citep{bia06}. This type of emission from ``classical'' Seyfert 2
galaxies has always been seen in the spectra of objects well studied
so far. However, this sample is dominated by optically selected
Seyfert 2 galaxies, which require a scattered component to be selected.

The scattered fraction is proportional to both the solid angle of the
scattering region as viewed from the nucleus, $\Omega_{\rm scat}$, and
the scattering optical depth, $\tau_{\rm scat}$. Hence, the observed
small scattered fraction means small $\Omega_{\rm scat}$ and/or small
$\tau_{\rm scat}$ i.e., deficiency of gas in the circumnuclear
environment, for some unknown reason.

The first possibility, which we favor as a more plausible case,
indicates that these AGNs are buried in a very geometrically-thick
obscuring torus. Assuming that the typical scattering fraction of 3\%
corresponds to the effective torus half-opening angle (see
\citealt{lev02} for definition) $\theta$ of 45 degree, our results
($<0.5\%$) indicate $\theta \simlt 20$ degree. In the case of \swsix ,
the small absorption for the reflection component, which probably comes
from the inner wall of the torus, suggests that we are seeing this
source in a rather face-on geometry. Indeed, applying the formalism of
\citet{lev02} to the observed EW of the iron-K line, we infer that the
inclination angle with respect to the axis of the disk, $i$, is smaller
than 40 degrees if $\theta < 20$ degree. For \swone , the presence of a
high column density for the reflection component implies a more edge-on
geometry than in \swsix. The observed EW of $0.20\pm0.05$ keV can be
explained if the torus is patchy or has a geometrical structure such
that the line-of-sight column density is much smaller than that in the
disk plane.

We infer that this type of buried AGNs is a significant fraction of the
whole AGN population, although an accurate estimate of this fraction is
difficult at present due to the small number statistics\footnote{We note
that a similar object has also been found by \citet{com07} with the
\suzaku\ follow-up of hard X-ray ($>10$ keV) selected AGNs.}. The
observed fraction of heavily obscured AGNs with log \nh\ $>23.5$ is
about 25\% among the hard X-ray ($E>$15 keV) selected AGNs
\citep{mar05}. The true number density of obscured AGNs could be much
larger, however. If we saw the same system of our targets at much larger
inclination angles ($i \gg 40$ degree), the observed flux of the
transmitted component would be much fainter even in hard X-rays due to
the effects of repeated Compton scatterings \citep{wil99}. Our results
imply that there must be a large number of yet unrecognized, Compton
thick AGNs in the local universe, which are likely to be missed even in
the \swift\ and \integral\ surveys.

The existence of AGNs with a geometrically thick torus was predicted by
\citet{fab98}, where the extreme obscuration was postulated to be caused
by a nuclear starburst. Using the $60 \mu$m and $100 \mu$m fluxes
measured by {\it Infrared Astronomical Satellite (IRAS)}, we obtain the
far infrared luminosity (defined by \citealt{dav92}) of $L_{\rm FIR} =
4.4\times10^{43}$ \erg\ and $7.1\times10^{43}$ \erg , and hence the
ratio between the 2--10 keV to far infrared luminosities of $L_{\rm
2-10}/L_{\rm FIR} \approx 0.02$ and $\approx 0.5$ for \swsix\ and \swone
, respectively. While the result of \swone\ is consistent with those of
the 2--10 keV selected AGNs in the local universe \citep{pic82} within
the scatter, the small $L_{\rm 2-10}/L_{\rm FIR}$ ratio of \swsix\
indicates a possibly significant starburst activity. However, this is
not supported by the optical spectrum of this object, which apparently
shows no evidence for a significant amount of star formation. The reason
behind the difference between the two sources is unclear.

By using the unique combination of the \swift\ BAT survey and the
\suzaku\ broad band spectral capabilities, we are discovering a new
type of AGN with an extremely small scattering fraction. This class of
object is most likely to contain a buried AGN in a very
geometrically-thick torus. This population was missed in previous
surveys, demonstrating the power of hard X-ray ($>$10 keV) surveys to
advance our global understanding of the whole AGN population. In
particular, we predict that the objects should have fainter [O~III]
emission luminosity relative to the hard X-ray luminosity compared
with classical Seyfert 2 galaxies because much less of the nuclear
flux ``leaks'' out to ionize the narrow line gas. As shown above, the
optical spectrum of \swsix\ is consistent with this prediction. This
study is particularly important since the existence of numerous such
objects would make surveys that rely on the [O~III] emission
incomplete by missing many of buried AGNs and incorrectly estimating
the true AGN luminosity.

\acknowledgments

We thank the members of the \suzaku\ team for calibration efforts of
the instruments, in particular Motohide Kokubun and Yasushi Fukazawa
for their useful advice regarding the background of the HXD. We would
also like to thank the anonymous referee for providing helpful
suggestions to improve this paper. Part of this work was financially
supported by Grants-in-Aid for Scientific Research 17740121 and
17740124, and by the Grant-in-Aid for the 21st Century COE ``Center
for Diversity and Universality in Physics'' from the Ministry of
Education, Culture, Sports, Science and Technology (MEXT) of Japan.

\end{document}